\documentclass[twocolumn,aps,prb,tightenlines,amsmath,amssymb]{revtex4}
\usepackage{graphicx}
\usepackage{amssymb}
\usepackage{dcolumn}
\usepackage{amsmath}
\usepackage{bm}

\usepackage{colordvi}

\begin{document}

\title{Comment on ``Density dependence of electron-spin polarization
  and relaxation in intrinsic GaAs at room temperature''}

\author{J. H. Jiang}
\author{M. W. Wu}
\thanks{Author to whom correspondence should be addressed}
\email{mwwu@ustc.edu.cn.}
\affiliation{Hefei National Laboratory for Physical Sciences at
Microscale and Department of Physics,
University of Science and Technology of China, Hefei,
Anhui, 230026, China}

\date{\today}

\begin{abstract}
We comment on the conclusion by Teng et al. [J. Phys. D: Appl. Phys. {\bf 42},
135111 (2009)] that the Bir-Aronov-Pikus mechanism is more important
than the D'yakonov-Perel' mechanism at high carrier density in
intrinsic bulk GaAs. We point out that the spin relaxation is solely
from the D'yakonov-Perel' mechanism.
\end{abstract}

\maketitle

Recently Teng et al. measured the density dependence of electron spin
relaxation time in intrinsic bulk GaAs at room temperature.\cite{Teng} They
found that the electron spin relaxation time decreases with increasing
carrier density in the carrier density regime
$10^{17}<N_{c}<2\times10^{18}$~cm$^{-3}$. Using the wrong formulae,
they found that the D'yakonov-Perel' (DP) spin relaxation time
increases with increasing carrier density. As the Bir-Aronov-Pikus
(BAP) spin relaxation time decreases with the carrier density, they
concluded that the BAP mechanism is more important than the DP
mechanism at high carrier density.

Their conclusion can not be correct. As shown in our recent
paper\cite{jiang} that the BAP mechanism is less important than
the DP mechanism in almost all the intrinsic bulk III-V semiconductors.
In fact, Teng et al. obtained such incorrect conclusion because they
used wrong statistics: they applied the Boltzmann statistics to a high
carrier density regime where $E_{\rm F}$ is comparable with or larger
than $k_{\rm B}T$. Then they obtained the increase of the DP spin
relaxation time with elevating carrier density as using the Boltzmann
statistics, the inhomogeneous broadening of the spin-orbit field
$\langle \Omega_{\bf k}^2\rangle$ does not change with carrier
density. Furthermore, they used an incorrect formula that the
electron-electron Coulomb scattering rate increases with carrier
density as $1/\tau_p^{ee}\sim N_c^{0.3}$ for such high density in the
experiment (The correct one can be found in
Ref.~\onlinecite{jiang}). Therefore, they obtained that the DP spin
relaxation time $\tau_{\rm DP}\sim 1/(\langle \Omega_{\bf
  k}^2\rangle\tau_p)\sim N_c^{0.3}$, which increases with increasing
carrier density.

\begin{figure}[htb]
\includegraphics[height=5.cm]{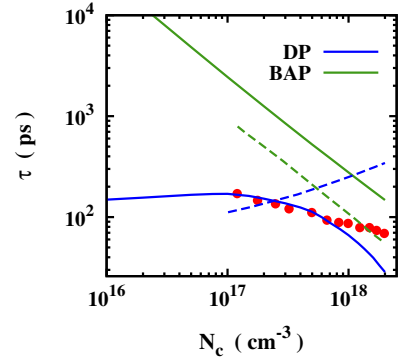}
\caption{(Color online) Carrier density $N_c$ dependence of spin
  relaxation time in intrinsic bulk GaAs at room temperature. The red dots
  represent the experimental results by Teng et al., \cite{Teng} the
  blue (green) curves are the spin relaxation time limited by the DP
  (BAP) mechanism. The solid curves are calculated via the fully
  microscopic kinetic spin Bloch equation approach, whereas the dashed
  ones are from the calculation in the paper of Teng et al.\cite{Teng}}
\end{figure}

In fact in the {\em non-degenerate} (low carrier density) regime the DP spin
relaxation time does increase with carrier density.\cite{jiang} However, for the high 
carrier density in the experiment, the electron system is actually in
{\em degenerate} regime (e.g., at $N_c=10^{18}$~cm$^{-3}$, $E_{\rm
  F}\simeq 2k_{\rm B}T$). In degenerate regime, the electron-electron
and electron-hole scatterings {\em decrease} with
increasing carrier density.\cite{jiang} Furthermore, the inhomogeneous
broadening $\langle \Omega_{\bf k}^2\rangle$ {\em increases} with
carrier density, $\langle \Omega_{\bf k}^2\rangle \sim k_{\rm F}^6\sim N_c^2$.
Therefore the DP spin relaxation time 
$\tau_{\rm DP}\sim 1/(\langle \Omega_{\bf k}^2\rangle\tau_p)$ 
decreases rapidly with increasing carrier density
in degenerate regime. There is {\em no way} that the DP spin
relaxation time can increase with carrier density at such high carrier
density in the experiment of Teng et al.\cite{Teng}

To check their results, we further calculate the spin relaxation time
via the fully microscopic kinetic spin Bloch equation
approach\cite{ksbe,jiang} which has been applied to many situations
with good agreement with experiments.\cite{exp} This many-body
approach includes all the relevant scatterings such as the
electron-impurity, electron-phonon, electron-electron Coulomb,
electron-hole Coulomb and electron-hole exchange scatterings 
explicitly.\cite{ksbe,jiang} Our results are plotted in Fig.~1 as
solid curves. For comparison the results of Teng et al. are plotted as
dashed curves. It is noted that the DP spin relaxation time from the
fully microscopic approach decreases with carrier density for
$N_c>10^{17}$~cm$^{-3}$, whereas it increases with increasing carrier
density at lower densities. This further confirms the above conclusion that the DP
spin relaxation time can only {\em decrease} with carrier density in
the experiment of Teng et al.

To further check the BAP spin relaxation time in the paper of
Teng et al.,\cite{Teng} we calculate the same quantity via the fully microscopic
kinetic spin Bloch equation approach.\cite{jiang} We find that Teng et
al. also overestimated the BAP mechanism (see Fig.~1). The possible
reason is that Teng et al. used larger electron-hole exchange
interaction constants to fit their experimental results. However,
these constants have been measured accurately and can be found in
standard handbooks such as {\em Landolt-B\"ornstein}.\cite{para} In
our fully microscopic calculation, all the material parameters are
taken from {\em Landolt-B\"ornstein}.\cite{para} 

From the results in Fig.~1, one can conclude that the experimental
results can not be explained via the BAP mechanism, as the BAP spin
relaxation time is much larger than the measured one. We then fit
the experimental results via the DP mechanism. For the DP spin relaxation,
there is only one free parameter, i.e., the Dresselhaus spin-orbit 
coupling constant $\gamma_{\rm D}$ which has not been unambiguously
determined by experiment or theory. With this parameter (which
actually scales the DP spin relaxation time as $\tau_{\rm DP}\propto
\gamma_{\rm D}^{-2}$), we fitted the experimental results. A best
fitting at low carrier density gives 
$\gamma_{\rm D}=7.6$~eV$\cdot$\AA$^{3}$. This value is close to the
value fitted from other experiment 
($\gamma_{\rm D}=8.2$~eV$\cdot$\AA$^{3}$)\cite{jiang} and that from 
recent {\it ab initio} calculation with GW approximation 
($\gamma_{\rm D}=8.5$~eV$\cdot$\AA$^{3}$).\cite{chantis} The
calculation agrees well with the experimental results for carrier densities up to
$10^{18}$~cm$^{-3}$. The discrepancy at high carrier density may come
from overestimation of the carrier density in the experiment and/or
the hot-electron effect due to optical excitation with excess 
carrier energy.

This work was supported by the Natural Science Foundation of China
under Grant No.~10725417. We thank T. S. Lai for helpful discussions.


\begin{thebibliography}{0}
\bibitem{Teng} L. H. Teng, K. Chen, J. H. Wen, W. Z. Lin, and
  T. S. Lai, J. Phys. D: Appl. Phys. {\bf 42}, 135111 (2009).
\bibitem{jiang} J. H. Jiang and M. W. Wu, Phys. Rev. B {\bf 79},
  125206 (2009).
\bibitem{ksbe} M. W. Wu and C. Z. Ning, Eur. Phys. J. B {\bf 18}, 373
  (2000); M. W. Wu and H. Metiu, Phys. Rev. B {\bf 61}, 2945
  (2000); M. Q. Weng, M. W. Wu, and L. Jiang, Phys. Rev. B
  {\bf 69}, 245320 (2004); M. Q. Weng and M. W. Wu, Phys. Rev. B {\bf
    68}, 075312 (2003).
\bibitem{exp} J. Zhou, J. L. Cheng, and M. W. Wu, Phys. Rev. B {\bf
    75}, 045305 (2007); D. Stich, J. Zhou, T. Korn, R. Schulz, D. Schuh,
  W. Wegscheider, M. W. Wu, and C. Sch\"uller, Phys. Rev. Lett. {\bf
    98}, 176401 (2007); Phys. Rev. B {\bf 76}, 205301
  (2007); D. Stich, J. H. Jiang, T. Korn, R. Schulz, D. Schuh,
  W. Wegscheider, M. W. Wu, and C. Sch\"uller, Phys. Rev. B {\bf 76},
  073309 (2007); L. H. Teng, P. Zhang, T. S. Lai, and M. W. Wu,
  Europhys. Lett. {\bf 84}, 27006 (2008); K. Shen, Chin. Phys. Lett. {\bf 26}, 067201 (2009).
\bibitem{para} {\it Semiconductors}, Landolt-B\"ornstein, New Series,
Vol.\ 17a, ed. by O. Madelung (Springer-Verlag, Berlin, 1987).
\bibitem{chantis} A. N. Chantis, M. van Schilfgaarde, and T. Kotani,
  Phys. Rev. Lett. {\bf 96}, 086405 (2006)

\end{thebibliography}
\end{document}